\documentclass[journal]{IEEEtran}
\usepackage{color}
\usepackage{graphicx}
\usepackage[dvips]{psfrag}
\usepackage{psfrag}
\usepackage{subfigure}
\usepackage{stfloats}
\usepackage{amsfonts}
\usepackage{amsbsy}
\usepackage{amsmath}
\usepackage{amssymb}
\usepackage{latexsym}
\usepackage{url}
\usepackage{multirow}
\usepackage{rotating}
\usepackage{cite}
\usepackage{dsfont}
\usepackage{widetext}

\begin{document}

\title{A Fully Distributed Approach for Plug-in Electric Vehicle Charging}

 \author{\IEEEauthorblockN{Javad Mohammadi\IEEEauthorrefmark{1},
 Marina Gonz\'alez Vay\'a\IEEEauthorrefmark{2},
 Soummya Kar\IEEEauthorrefmark{1},
 Gabriela Hug\IEEEauthorrefmark{2}}\\
 \IEEEauthorblockA{\IEEEauthorrefmark{1} Department of Electrical and Computer Engineering\\
 Carnegie Mellon University,
 Pittsburgh, PA, 15213\\ {jmohamma, soummyak}@andrew.cmu.edu\\}
 \IEEEauthorblockA{\IEEEauthorrefmark{2} Power Systems Laboratory\\ ETH Zurich,
 Zurich\\ {gonzalez, hug}@eeh.ee.ethz.ch}}

\maketitle

\begin{abstract}
Plug-in electric vehicles (PEVs) are considered as flexible loads since their charging schedules can be shifted over the course of a day without impacting drivers' mobility. This property can be exploited to reduce charging costs and adverse network impacts. The increasing number of PEVs makes the use of distributed charging coordinating strategies preferable to centralized ones. In this paper, we propose an agent-based method which enables a fully distributed solution of the PEVs' Coordinated Charging (PEV-CC) problem. This problem aims at coordinating the charging schedules of a fleet of PEVs to minimize costs of serving demand subject to individual PEV constraints originating from battery limitations and charging infrastructure characteristics. In our proposed approach, each PEV's charging station is considered as an agent that is equipped with communication and computation capabilities. Our multi-agent approach is an iterative procedure which finds a distributed solution for the first order optimality conditions of the underlying optimization problem through local computations and limited information exchange with neighboring agents. In particular, the updates for each agent incorporate local information such as the Lagrange multipliers, as well as enforcing the local PEV's constraints as local innovation terms. Finally, the performance of our proposed algorithm is evaluated on a fleet of 100 PEVs as a test case, and the results are compared with the centralized solution of the PEV-CC problem.

\end{abstract}

\vspace{0.15cm}
\begin{IEEEkeywords}
Distributed Updates, Plug-in Electric Vehicles, \textit{consensus+innovation}-based Approach, Coordinated Charging, Optimality Conditions
\end{IEEEkeywords}

\IEEEpeerreviewmaketitle

\section*{Nomenclature}

\begin{table}[!h]
 \centering\normalsize
  \begin{tabular}{  p{1.7cm}  p{6.5cm} }
  $\mathbf{x_{v}}$& Charging power schedule of PEV $v$ over a given time horizon $[0,T]$, $\mathbf{x_v}\in \mathbb{R}^{T\times 1}$\\
  $\mathbf{L}$& Aggregated load of PEVs over a given time horizon $[0,T]$, $\mathbf{L}\in \mathbb{R}^{T\times 1} $\\
  $\Omega_v$& Set of neighboring charging stations connected to charging station $v$\\
  $c_1$, $c_2$&Cost function parameters, $c_1\in \mathbb{R}$, $c_2\in \mathbb{R}^{1\times T}$\\
  $A$, $b_v$  & Matrix and vector defining the energy constraints of an individual PEV\\
  $\underline{x}_{v}$, $\overline{x}_{v}$ & Upper and lower bounds defining the power constraints of an individual PEV\\
  $\lambda$, $\mu$& Lagrangian multipliers associated with equality and inequality constraints\\
 \end{tabular}
 \label{table:Vars}
 \end{table}

\begin{table}[!h]
 \centering\normalsize
  \begin{tabular}{  p{1.7cm}  p{6.5cm} }
  $\mu_{-}$, $\mu_{+}$&Lagrangian multipliers associated with decision variables' upper and lower bounds\\
  $V$&Total number of PEVs\\
 \end{tabular}
 \label{table:Vars}
 \end{table}
\section{Introduction}
\subsection{Motivation}
It has been argued that the uncoordinated charging of plug-in electric vehicles (PEVs) can be potentially challenging for the power network \cite{Verzijlbergh12}. However, PEV electricity demand is very flexible and can therefore be adjusted in time to reduce the costs of providing the charging demand and to avoid undesirable effects on the network.


Different types of approaches to the charging coordination problem, mainly centralized, and decentralized\footnote{In the literature, the term ``decentralized" is used to refer to approaches that do not rely on communication, i.e., where charging decisions are taken by PEVs solely based on local information. It can also refer to approaches where PEVs take their own charging scheduling decisions based on information shared with a central coordinator.}/distributed approaches, have been proposed in the literature. In most approaches, an entity called ``aggregator" is considered as an intermediary agent between the PEVs and other power system entities, such as network operators or energy providers. This aggregator can play a more or less passive role depending on the type of approach considered.

Centralized approaches \cite{Bessa2012,Vagropoulos13,Gonzalez14a,amini2013probabilistic,hu2014coordinated,rotering2011optimal,amini2014allocation} face mainly two challenges: they may require the communication of sensitive information (arrival and departure times, energy requirements) from PEVs to an aggregator, and  they are typically not scalable. \textcolor{black}{On the other hand, sharing control responsibilities among entities could decrease the complexity of the charging coordination problem, and hence, provides a scalable solution approach compared with centralized control strategies.}

\subsection{Related Work}

Most communication-based decentralized approaches introduced so far require the exchange of information with an aggregator, which acts as a coordinating agent \cite{Ma2013,Gan13,ParColGraLyg14,Rivera2013,GonAndBoy14,rahbari2014cooperative}. The information exchanged is however not sensitive (typically the charging schedule and dual variables). The approaches in  \cite{Ma2013,ParColGraLyg14} consider non-cooperative agents and are based on mean field game theory, whereas the approaches in \cite{Gan13,Rivera2013,GonAndBoy14,rahbari2014cooperative} consider cooperative agents. The charging optimization problem is decomposed using the Alternating Direction Method of Multipliers in \cite{Rivera2013,GonAndBoy14}. The decentralized approaches mentioned above require each PEV to communicate with a central agent and are therefore less robust towards failure than peer-to-peer based distributed schemes.

Recently, consensus-based  approaches \cite{olfati2007consensus} have been used to provide distributed control schemes for applications in electric power systems such as solving optimal power management problems \cite{asr2013consensus,kar2014distributed,kargarian2014optimal}, the Economic Dispatch problem \cite{zhang2012convergence,kar2012distributed2,binetti2014distributed,yang_consensus_2013} and Optimal Power Flow problems \cite{Mohammadi_distributedOPF_2014(1)}. A neighborhood consensus potential \cite{dimakis2010gossip,olfati2007consensus} in the iterative update procedure ensures that entities reach an agreement on a common variable, usually corresponding to electricity price in the aforementioned problems.

In \cite{rahbari2014cooperative}, a consensus-based method to coordinate PEV charging is proposed. However, it requires one of the agents to access information on the total charging demand. Moreover, \cite{xu2015optimal} proposes a consensus-based distributed charging rate control strategy for a PEV fleet to minimize total charging power loss, which overlooks PEV's limitations.

This paper presents a fully distributed multi-agent method to solve the PEVs' Coordinated Charging (PEV-CC) problem. Our method denoted by $\mathcal{CI-PEVCC}$, i.e., \textit{consensus+innovation} based PEV Coordinated Charging, solves the first order optimality conditions of the original problem. The \textit{consensus} update term of the algorithm facilitates the agreement on an incremental price for the energy provided. The \textit{innovation} term ensures that total consumption of a PEV fleet yields the minimum cost while individual PEV's local constraints are satisfied.

\begin{figure}[t]
 \centering
\includegraphics[width=0.32\textwidth]{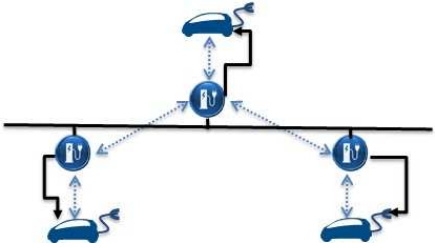}
 \caption{Proposed distributed PEVs' coordinated charging scheme (the dashed line represents data exchange, and the solid line represents charing power)
 }
 \label{PEV-CC-tst}
\end{figure}

\subsection{Contribution}
Our proposed $\mathcal{CI-PEVCC}$ inherently differs from decomposition theory-based methods in multiple ways: methodologically, our method directly solves the first order optimality conditions of the original problem. Hence, it technically reduces the original optimization problem to finding solutions for a coupled system of linear equations with geometric constraints in a fully distributed manner. While $\mathcal{CI-PEVCC}$ allows for a fully distributed calculation of the solution to the PEV-CC problem, it does not require communication of sensitive information among PEVs. Also, the inter-PV communication graph is assumed to be connected, but could be arbitrary (for example, very sparse) otherwise. Note, the communication structure can be defined arbitrarily, and increasing the number of communication links improves the information propagation speed, hence, increases convergence speed. Finally $\mathcal{CI-PEVCC}$ is a scalable solution to the PEV coordination problem since it distributes the computation and communication burden among PEVs. Figure~\ref{PEV-CC-tst} illustrates the communication structure.
\subsection{Paper Organization}
The rest of the paper is structured as follows: the PEVs'  Coordinated Charging problem formulation is given in Section II. The $\mathcal{CI-PEVCC}$
algorithm is presented in Section III. Section IV describes the test case specifications and presents simulation results. Finally, Section V concludes the paper.

\section{Problem Formulation}
\subsection{PEV's Coordinated Charging}

We consider a charging problem where a group of PEVs minimizes a common cost function, which is quadratic with respect to their aggregated load $\mathbf{L}$ over horizon $T$ \cite{GonAndBoy14}, subject to power and energy constraints, i.e.,
\begingroup
\allowdisplaybreaks
\begin{align}\label{GeneralPHEVcharging}
\mathrm{minimize}_{\mathbf{x}_{v},\mathbf{L}}&~~\mathbf{L}^\top \cdot c_1 \cdot \mathbf{L}+c_2^\top \cdot \mathbf{L}\\
\textrm{s.t.} ~~\mathbf{L}&=\sum_{v\in V}\mathbf{x}_{v},\label{loadEq_const}\\
&A\cdot  \mathbf{x}_{v}\leq b_{v},~~~\forall v\in\{1,\cdots ,V\} \label{en_const},\\
\underline{x}_{v}&\leq \mathbf{x}_{v} \leq \overline{x}_{v},~~~\forall v\in\{1,\cdots ,V\}. \label{pow_const}
\end{align}
\endgroup
The parameters $c_1$ and $c_2$ are functions of the total predicted inelastic load (load other than PEV load). Specifically, as in \cite{GonAndBoy14}, we assume that the goal is to minimize the costs of serving both the PEV and the inelastic load, and that these costs are a quadratic function of the sum of PEV and inelastic load.\footnote{The cost of serving PEV demand $\mathbf{L}$ and inelastic demand $\mathbf{L}_\mathrm{in}$ has the form $\widetilde{a}\mathbf{1}^\top(\mathbf{L}+\mathbf{L}_\mathrm{in})+\widetilde{b}(\mathbf{L}+\mathbf{L}_\mathrm{in})^\top(\mathbf{L}+\mathbf{L}_\mathrm{in})$, where $\widetilde{a}$ and $\widetilde{b}$ are scalars. Minimizing this cost function is equivalent to minimizing \eqref{GeneralPHEVcharging} with appropriate choices for $c_1$ and $c_2$.}

Constraints \eqref{en_const} represent the energy constraints, i.e., constraints on the cumulative demand of an individual load. These constraints ensure that the upper and lower State-Of-Charge (SOC) bounds of any PEV's battery are not violated, given the information on connection times and trip energy consumption. These constraints also ensure that the energy in the battery at the beginning and end of the optimization horizon are equal, to avoid battery depletion due to the cost minimization objective.

Note, here we solve a day-ahead optimization problem over the period of one day, within which vehicles can plug in and out several times. Therefore, in our problem, PEVs do not only optimize for a single parking instance, but optimize their schedules over a whole day. To be more specific, we do not set specific SOC requirements at the end of each parking period but ensure that, first of all the upper and lower SOC bounds are not violated. Secondly, the vehicle recharges the energy it consumes throughout the day within the 24 hours. This is equivalent to ensuring that the SOC at the beginning and end of the optimization period, i.e., one day, should be the same.

Equation \eqref{en_const} is an abstract representation of the energy limitation of PEVs' batteries. To discuss the derivation of  this constraint in detail, we assume that the initial energy content of the battery is given by $E_{0,v}$. Moreover, the charging efficiency and time step duration are denoted by $\eta_v$ and $\Delta t$, repectively. Finally, the energy consumption at each time step $t$ is represented by $E_{\text{cons},v}(t)$. Then, the energy content of a PEV's battery at a given time step is determined by
\begin{equation*}
E_{v}(t)=E_{0,v}+\eta_v\Delta t \sum_{\tau=1}^t x_{v}(\tau)-\sum_{\tau=1}^t E_{\text{cons},v}(\tau).
\end{equation*}
The upper and lower bounds on the energy content are given by the battery capacity $C_v$ and the minimum state of charge $\underline{\mathrm{SOC}}_{v}$
requirements, resulting in
\begin{equation}
 \underline{\mathrm{SOC}}_{v}\leq \frac{E_{v}(t)}{C_v}\leq 1.
\end{equation}

 Moreover, to avoid myopic behavior we force the energy content at the end of the time horizon to be equal to the initial energy content $E_{0,v}=E_{v}(T)$. Otherwise, due to cost minimization, the vehicle would tend to deplete the battery at the end of the optimization horizon. Finally, for the energy constraints written in the compact form
\eqref{en_const}, the right hand side is a function of
\begin{equation*}
b_{v}=h(C_v,\underline{\mathrm{SOC}}_{v},\eta_v,E_{\text{cons},v},E_{0,v}).
\end{equation*}
Further details on the derivation of $b$ are presented in \cite{gonzalez2012centralized}.

The power constraints \eqref{pow_const} set upper and lower bounds on the charging power. Since we consider unidirectional charging only in this paper, the lower bound on the charging
power is set to zero ($\underline{x}_{v} = 0$). The upper bound $\overline{x}_{v}$
is zero during the time steps when the vehicle is not connected,
and equal to the maximum charging rate of the charging infrastructure or the battery, $\overline{P}_{v}$, when the vehicle is
connected,
\[
   \overline{x}_{v}(t)=
\begin{cases}
    \overline{P}_{v},& c_{v}(t)=1\\
    0,              & c_{v}(t)=0.
\end{cases}
\]
where $c_{v}(t)$ is a binary variable describing the connection status at time step $t$. Therefore, the value of the upper bound is affected
by the timing of PEV trips. It should be noted that using this binary variable further means that a PEV can plug in and out several times during the optimization horizon. In summary, \eqref{en_const} and \eqref{pow_const} are local constraints, i.e., merely involve variables of an individual PEV, while  \eqref{loadEq_const} is the global constraint, includes variables from all PEVs.

\subsection{Optimality Conditions}
The Lagrangian function for the aforementioned optimization problem is given by
\begin{eqnarray}
\mathfrak{L}  &=&  \mathbf{L}^\top \cdot c_1 \cdot \mathbf{L}+c_2 \cdot \mathbf{L} \nonumber\\
&&  + \lambda^\top\cdot\left( -\mathbf{L}+\sum_{v\in V}\mathbf{x}_{v}\right)\nonumber\\
&& +\mu^\top_{v}\cdot\left( A\cdot  \mathbf{x}_{v}- b_{v}\right)\nonumber\\
&& +  \mu^\top_{v,-} \cdot\left(\underline{x}_{v}- \mathbf{x}_{v}\right)+  \mu^\top_{v,+}\cdot \left(\mathbf{x}_{v}-\overline{x}_{v} \right),
\end{eqnarray}
where $\lambda$ and $\mu$s are Lagrange multipliers. The corresponding first order optimality conditions are derived as follows,
\begingroup
\allowdisplaybreaks
\begin{alignat}{6}
\frac{\partial \mathfrak{L}}{\partial \mathbf{L}} &=& 2c_1 \cdot \mathbf{L}+c_2 - \lambda&=&0,\label{KKT1}\\
\frac{\partial \mathfrak{L}}{\partial \mathbf{x}_{v}} &=& \lambda+A^\top\cdot \mu+\mu_{+}-\mu_{-}&=&0,\label{KKT2}\\
\frac{\partial \mathfrak{L}}{\partial \lambda} &=& -\mathbf{L}+\sum_{v\in V}\mathbf{x}_{v}&=&0,\label{KKT3}\\
\frac{\partial \mathfrak{L}}{\partial \mu_{v}} &=& A\cdot  \mathbf{x}_{v}- b_{v}&\leq&0,\label{KKT4}\\
\frac{\partial \mathfrak{L}}{\partial \mu_{v,+}} &=&  \mathbf{x}_{v}- \overline{x}_{v}&\leq&0,\label{KKT5}\\
\frac{\partial \mathfrak{L}}{\partial \mu_{v,-}} &=&  -\mathbf{x}_{v}+ \underline{x}_{v}&\leq&0,\label{KKT6}
\end{alignat}
\endgroup
for all $v\in\{1,\ldots,V\}$ plus the complementary slackness conditions for the inequality constraints, i.e,
\begingroup
\allowdisplaybreaks
\begin{alignat}{6}
\mu_v\cdot \left(A\cdot  \mathbf{x}_{v}- b_{v}\right)&=&0,\\
\mu_{v,+}\cdot \left(\mathbf{x}_{v}- \overline{x}_{v}\right)&=&0,\\
\mu_{v,-}\cdot\left(-\mathbf{x}_{v}+ \underline{x}_{v}\right)&=&0,
\end{alignat}
\endgroup
and additionally we impose positivity constraints on the $\mu_{v}$, $\mu_{v,+}$, and $\mu_{v,-}$'s. Consequently, in order to find a solution to the PEV-CC problem, the above system of equations needs to be solved. Since the discussed problem is convex and also fulfills the strong duality conditions, any solution that satisfies all of the discussed first order optimality conditions is the optimal solution of the PEV-CC problem. Here, we assume that the primal optimization problem is strictly feasible.

\section{Distributed Scheduling}
\subsection{Distributed Decision-Making}
In this section, we present a brief review of the generic \textit{consensus+innovation} approach for solving collaborative distributed decision-making processes (see \cite{kar2012distributedparameter}). In the \textit{consensus+ innovation} setup each agent or decision maker performs local information processing and communication with neighboring agents to optimize a global decision-making task. The underlying assumption is that each agent has access to merely local information and the inter-agent data exchange is limited to a sparse communication graph.

Here, we focus on the application of a \textit{consensus+innovation} method to solve the distributed restricted agreement problem, i.e., reaching an agreement between $J$ agents on a common value $\mathbf{z}$ which satisfies the following global restriction
\begin{equation}
g(\mathbf{z})=\sum_{j=1}^{J}d_j(\mathbf{z})=\mathbf{Z},\label{globalCons}
\end{equation}
where, $d_j(.)$ is a certain real-valued function. This global constraint is also subject to local constraints of each agent $j$, e.g., upper and lower bounds restricting values of function $d_j(.)$.

Note, each agent $j$ is only aware of local information, i.e., its own information and information of neighboring agents ($\Omega_j$). In fact, $\Omega_j$ is a preassigned set which defines the interaction structure between agent $j$ and the rest of the agents. Under broad assumptions on $d_j(.)$ and the communication graph, an iterative \textit{consensus+innovation} type algorithm could be utilized to find a distributed solution for the discussed restricted agreement-type problem (for more details see \cite{hug2013consensus+}).

In the iterative process of the \textit{consensus+innovation} algorithm each agent holds and updates a local copy of $\mathbf{z}$ at each iteration $k$, denoted by $\mathbf{z}_j(k)$. The update of the local copy of the common variable follows the  format below,
\begin{eqnarray}
\mathbf{z}_j(k+1)&=&\mathbf{z}_j(k)-\beta_k \left(\sum_{w\in \Omega_j}(\mathbf{z}_j(k)-\mathbf{z}_w(k))\right)\nonumber\\
&-&\alpha_k \left(\widehat{\mathbf{Z}}_{j}-\widetilde{d}_j\right)\nonumber,\label{}
\end{eqnarray}
Here $\beta_k$ and $\alpha_k$  are tuning parameters. Also, $\widehat{\mathbf{Z}}_{j}$ is the estimation of agent $j$ of the global commitment. This estimation will be updated in each iteration based on the newly updated local information (This will be discussed in extensive details in the next section). Finally, the \textit{consensus+innovation} iterative procedure ensures that the updated local function lies in the predefined feasible region of $d_j(.)$,
\begin{equation}
\widetilde{d}_j=\mathbb{P}\left[d_j(\mathbf{z}_j(k))\right],\nonumber
\end{equation}
where $\mathbb{P}[.]$ denotes the projection operator onto the feasible space imposed by local constraints.

Typical conditions that ensure convergence, i.e., $\mathbf{z}_j(k) \rightarrow \mathbf{z}$ as $k \rightarrow \infty$ for all $j$ with $\mathbf{z}$ satisfying \eqref{globalCons} are as follows (see \cite{kar2012distributedparameter}):
\begin{enumerate}
\item The local functions $d_j(.)$s are sufficiently regular.
\item The inter-agent communication graph is connected.
\item The weight parameters $\alpha$ and $\beta$ are positive and satisfy the following conditions:
\begin{itemize}
\item The sequences ${\alpha_k}$ and ${\beta_k}$ are decaying, i.e., as $k \rightarrow \infty$, $\alpha_k \rightarrow 0$, $\beta_k \rightarrow 0$.
\item The weights are persistent, i.e.,
\begin{equation}
\sum_{k \geq 0} \alpha_k=\sum_{k \geq 0} \beta_k=\infty.\nonumber
\end{equation}
\item The innovation excitation decays at a faster rate than the consensus tuning parameter, i.e., $\beta_k/\alpha_k \rightarrow \infty$ as $k\rightarrow\infty$.
\end{itemize}
\end{enumerate}
Moreover, the \textit{consensus+innovation} approach has been shown to be resilient against data packet drops, random communication link failure and noisy information. Robustness of the \textit{consensus+innovation} approach in the context of distributed energy management is discussed in \cite{kar2012distributed2}.
\subsection{Distributed Updates}
In $\mathcal{CI-PEVCC}$, each PEV charging station is required to exchange information with all of its physically connected neighboring charging stations at each iteration. In our proposed approach, each PEV $v$ updates the variables $\mathbf{x}_{v}$, $\mathbf{L}_v$, $\mu_v$, and $\lambda_v$ which are directly associated with PEV $v$.
We denote the iteration counter by $k$ and the iterates by $\mathbb{X}_v(k)$, which include the variables associated with PEV $v$ at iteration k, i.e., $\mathbb{X}_v(k)=[\mathbf{x}_{v},\mathbf{L}_v, \mu_v, \lambda_v]$.

The Lagrange multipliers $\lambda_v$ are updated according to
\begin{eqnarray}
\lambda_v(k+1)&=&\lambda_v(k)-\beta_k \left(\sum_{w\in \Omega_v}(\lambda_v(k)-\lambda_w(k))\right)\nonumber\\
&-&\alpha_k \left(\frac{\mathbf{L}_v(k)}{V}-\mathbf{x}_v(k)\right)\label{LambdaUpdate},
\end{eqnarray}
where $\alpha_k$, $ \beta_k> 0$ are tuning parameters. The first term preserves the coupling between the Lagrange
multipliers, while ensuring that $\lambda$s are reaching \textit{consensus}.
The second term, referred to as \textit{innovation}, reflects the accuracy of PEV $v$'s estimation of the total load ($\mathbf{L}$). The update makes intuitive sense, e.g., if PEV $v$'s consumption ($\mathbf{x}_{v}$) is more than its estimated share of overall consumption ($\mathbf{L}_v(k)/V$), then the innovation term results in an increase in the value of $\lambda_v(k+1)$. Consequently, using \eqref{Lupdate} to update $\mathbf{L}_v$, PEV $v$'s estimation of overall load increases in the next iteration.

Knowing the value of the Lagrange multiplier $\lambda_v$, PEV $v$ could update its estimation of the total load ($\mathbf{L}_v$) by carrying out the following update,
\begin{eqnarray}
\mathbf{L}_v(k+1)&=&\mathbb{P}\left[\mathbf{L}_v(k)-\frac{1}{2c_1} \frac{\partial \mathfrak{L}}{\partial \mathbf{L}_v(k)} \right]_{[0,\infty)}\nonumber\\
&=&\mathbb{P}\left[\frac{\lambda_v(k)-c_2}{2c_1} \right]_{[0,\infty)},\label{Lupdate}
\end{eqnarray}
where $\mathbb{P}$ is the projection operator which projects the updated value of $\mathbf{L}_v$ into the feasible space, i.e., $[0,\infty)$. Note, $\mathbf{L}_v$ is the estimation of agent $v$ of the global PEVs' load. Also, our update structure requires all the agents (PEVs) to know cost function parameters, which is a reasonable assumption, since the electricity tariffs are generally predetermined (they need to be communicated once in advance and not in real-time).

The PEV's charging schedules are updated according to
\begin{eqnarray}
\mathbf{x}_{v}(k+1)\hspace{-.2cm}&=&\hspace{-.3cm}\mathbb{P}\left[\mathbf{x}_{v}(k)-\delta_k \frac{\partial \mathfrak{L}}{\partial \mathbf{x}_{v}(k)} \right]_{[\underline{x}_{v},\overline{x}_{v}]}\label{Xupdate}\\
&=&\hspace{-.3cm}\mathbb{P}\left[\mathbf{x}_{v}(k)-\delta_k\left(\lambda_v(k)+A^\top\cdot \mu_{v}(k) \right)\right]_{[\underline{x}_{v},\overline{x}_{v}]},\nonumber
\end{eqnarray}
with $\delta_k>0$ being another tuning parameter. The projection operator in this case projects $\mathbf{x}_{v}$ onto the feasible space determined by $[\underline{x}_{v},\overline{x}_{v}]$. In other words, if $\mathbf{x}_{v}$ violates any of the bounds, $\mathbb{P}$ sets the updated value equal to that bound. The \textit{innovation} term includes $\lambda_v$ and $\mu_v$. This makes intuitive sense because whenever $\lambda_v$, i.e., price of energy, increases, PEV $v$'s consumption ($\mathbf{x}_{v}$) decreases. Also, the presence of $\mu_v$ leads to proper adjustment of $\mathbf{x}_{v}$ to fulfill \eqref{KKT4}.

The update (\ref{Xupdate}) does not take into account $\mu_{v,+}$ and $\mu_{v,-}$ in (\ref{KKT1}), since these multipliers do not appear in any other constraint and the projection operator ensures the feasibility of the achieved update.

The update for $\mu_v$ is given by
\begin{eqnarray}
\mu_v(k+1)&=&\mathbb{P}\left[\mu_v(k)+\gamma_k \frac{\partial \mathfrak{L}}{\partial \mu_v(k)}\right]_{[0,\infty)}\label{MUupdate}\\
&=&\mathbb{P}\left[\mu_v(k)+\gamma_k \left(A\cdot\mathbf{x}_{v}(k)-b_v\right)\right]_{[0,\infty)},\nonumber
\end{eqnarray}
with $\gamma>0$ being a tuning parameter. This update uses the inequality (\ref{KKT4}). The projection operator $\mathbb{P}$ ensures the positivity of the Lagrangian multipliers $\mu$’s by setting the $\mu_v(k+1)$ equal to zero if the update (\ref{MUupdate}) results in negative values. Assuming that the current value for $\mathbf{x}_{v}$ satisfies (\ref{KKT4}), our proposed update yields a decreasing value for $\mu_v$ with a minimum value of zero due to the projection into the feasible space, i.e., $\mu_v\geq0$. If $A\mathbf{x}_{v}(k)>b_v$, then the value for $\mu_v$ may increase which will adjust $\mathbf{x}_{v}$ accordingly in the next update iteration (using \eqref{Xupdate}).

The pseudo code for the $\mathcal{CI-PEVCC}$ is given in Table \ref{Pseudo}. The stopping condition can be defined based on some user-defined criterion, e.g., the measurement of $\mathrm{rel}_\mathrm{obj}$ and $\mathrm{rel}_\mathrm{load}$, defined in Section \ref{sec_convmeas},  is simply for the sake of performance analysis. Also, $\mathcal{CI-PEVCC}$ only requires agents to exchange $\lambda_v$ during the course of iterations, which is not sensitive information.

It should be noted that all of the updates are purposely using the variables from the previous iteration which facilitates a parallel computation of all of the updates. The serial implementation improves convergence speed in terms of the number of iterations 
times but since the updates have to be done after each other the computation time per agent most likely increases.

Finally, $\mathcal{CI-PEVCC}$ is a fully distributed algorithm, i.e., each PEV is represented with one agent, and requires each agent to perform computations at each iteration. However, distributedness, i.e., the number of PEVs that are considered by a single agent, in $\mathcal{CI-PEVCC}$ could be defined based on the needs and characteristics of the agents. In this regard, \cite{mohammadi2015asynchronous} discusses the possibility of clustering entities to form an agent. Moreover, it presents an \textit{asynchronous} update scheme which could be used for multilevel implementation of a distributed algorithm. In a nutshell, as the number of agents decreases, the communication needs reduce while the computational burden of each agent increases.

\begin{table}[t]
\center
\caption{Pseudo code for the $\mathcal{CI-PEVCC}$ algorithm}\label{Pseudo}
\begin{tabular}{|l|}
\hline
Initialize tuning parameters\\
Initialize variables $\lambda_v,\mathbf{L}_{v},\mathbf{x}_{v},\mu_v$\\
\hspace*{1cm} While convergence criteria is not satisfied\\
\hspace*{1.5cm} for i=1:number of agents\\
\hspace*{2cm} Update $\lambda_v$ using (12)\\
\hspace*{2cm} Update $\mathbf{L}_{v}$ using (13)\\
\hspace*{2cm} Update $\mathbf{x}_{v}$ using (14)\\
\hspace*{2cm} Update $\mu_v$ using (15)\\
\hspace*{2cm} Communicate $\lambda_v$ to neighboring agents\\
\hspace*{1.5cm} end\\
\hspace*{1.5cm} measure $\mathrm{rel}_\mathrm{obj}$ and $\mathrm{rel}_\mathrm{load}$\\
\hspace*{1cm} end\\
\hline
\end{tabular}
\vspace{-.0cm}
\end{table}

\section{Simulation Results}
In this section, we provide a proof of concept by carrying out simulations.

\subsection{Simulation Setup}
We simulate a fleet of $V=100$ vehicles, with maximum charging power 11kW, charging efficiency 0.9, minimum state of charge 0.2, and battery size of either 16kWh or 24kWh. The driving pattern information is obtained from a transport simulation for Switzerland, with the tool MATSim \cite{Balmer2006}. These patterns are then translated into $b_v$ \cite{gonzalez2012centralized}. Therefore, $b_v$ represents the driving pattern, e.g., trips specifications, arrival and departure times, of PEV $v$.

Also, the load profile used in the simulation represents a typical winter load in the city of Zurich. It should be mentioned that the original load profile is scaled so that the total PEV charging demand constitutes 10\% of total demand. The optimization time horizon is one day, divided into 15 minutes time intervals. Therefore, $\mathbf{x}_v\in \mathbb{R}^{96\times 1}$, where $\mathbf{x}_v^1$ represents the charging of PEV $v$ at the end of first time interval (15 minutes).
 Also, the communication graph is considered as a path graph, i.e., each PEV exchanges information with two neighboring PEVs, except for the PEVs at the ends of the path which only communicate with one neighboring PEV, see Fig.\ \ref{PEV-CC-tst}.

\begin{figure}[b]
 \centering
\includegraphics[width=0.5\textwidth]{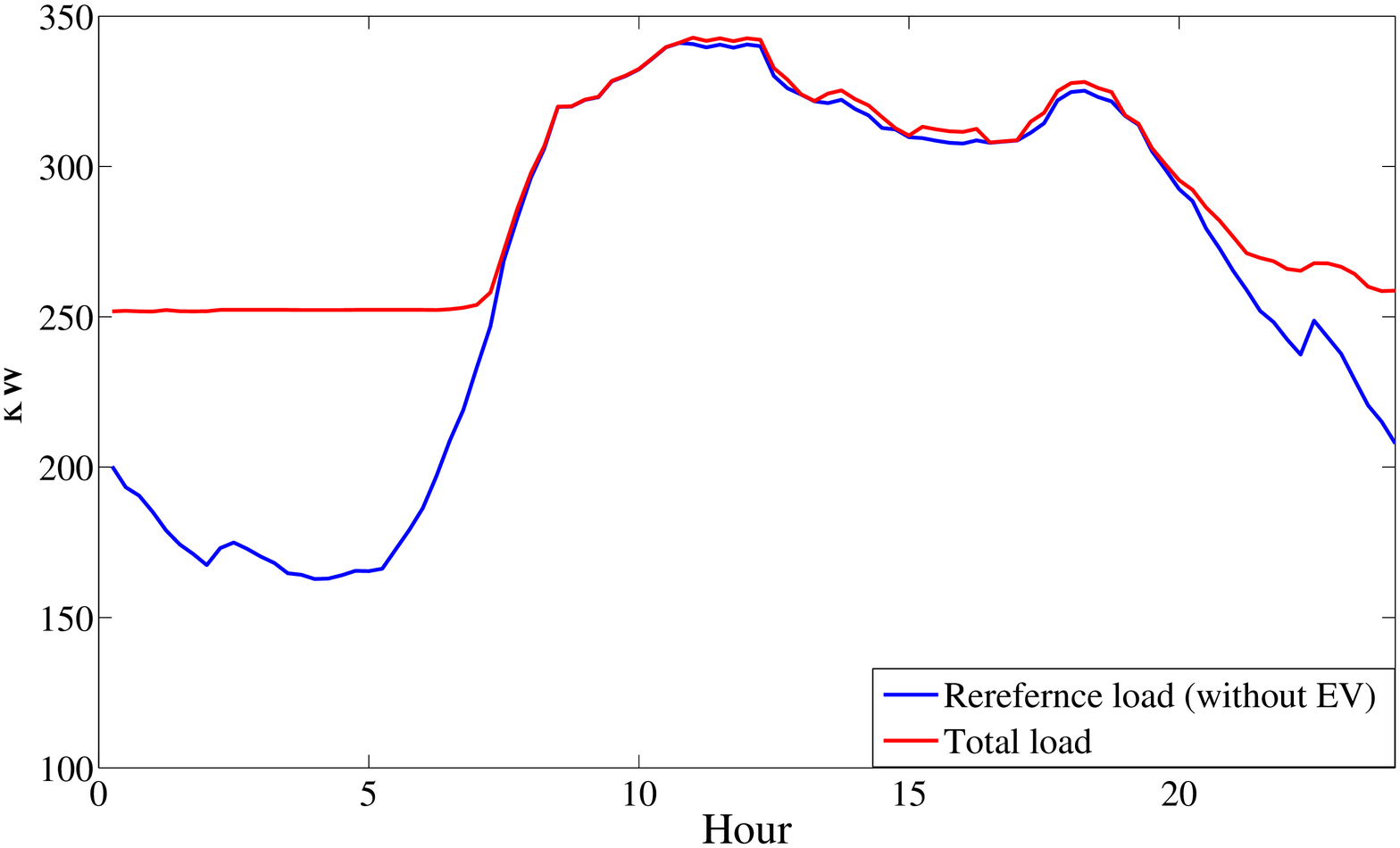}
 \caption{Total load profiles for the full 24h horizon.}\label{LOADprofile}
\end{figure}

\subsection{Convergence Measurements}\label{sec_convmeas}
In order to evaluate the performance of our $\mathcal{CI-PEVCC}$ approach, two measures are introduced in this paper. The relative distance of the objective function from the optimal value over the iterations is considered as the first measure,
\begin{equation}
\mathrm{rel}_\mathrm{obj}=\frac{\left | f-f^{*} \right |}{f^{*}},\label{rel}
\end{equation}
here $f^*$ is the optimal objective function value calculated by solving the centralized coordinated PEV charging problem. The value of $f^*$ is obtained from solving the centralized problem using the optimization package Tomlab in MATLAB environment.

Moreover, the second measure determines the relative distance of the total load at each iteration from the optimal value of overall PEV load calculated by solving the centralized problem ($\mathbf{L}^{*}$),
\begin{equation}
\mathrm{rel}_\mathrm{load}=\frac{\left | \sum_{t=1}^{T}\sum_{v=1}^{V}\mathbf{x}_{v}^{t}-\sum_{t=1}^{T}\mathbf{L}^{t,*} \right |}{\sum_{t=1}^{T}\mathbf{L}^{t,*}}.\label{rel}
\end{equation}

\subsection{Simulation Results}
The resulting load profile is depicted in Fig.\ \ref{LOADprofile}. Most charging demand is scheduled during the low-load hours of the night which contributes to  valley-filling. Also, charging partially takes place during the shoulder hours, i.e., the hours between the daily peaks. This further indicates that in these hours demand is not flexible enough to be shifted completely to the valley hours. Moreover, PEV demand merely has a negligible impact on the daily peak.

\subsection{Convergence}
\begin{figure}[t]
 \centering
\includegraphics[width=0.45\textwidth]{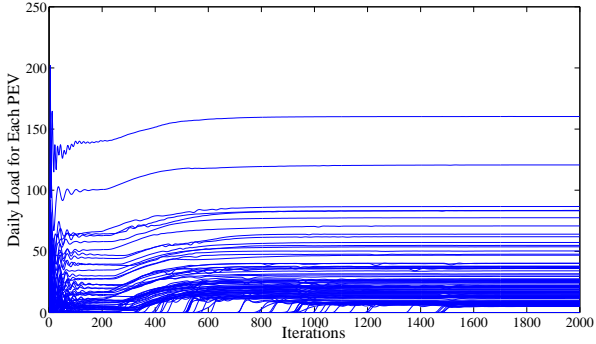}
 \caption{Daily charging load of each PEV.}
 \label{EVcharging}
\end{figure}

\begin{figure}[t]
 \centering
\includegraphics[width=0.5\textwidth]{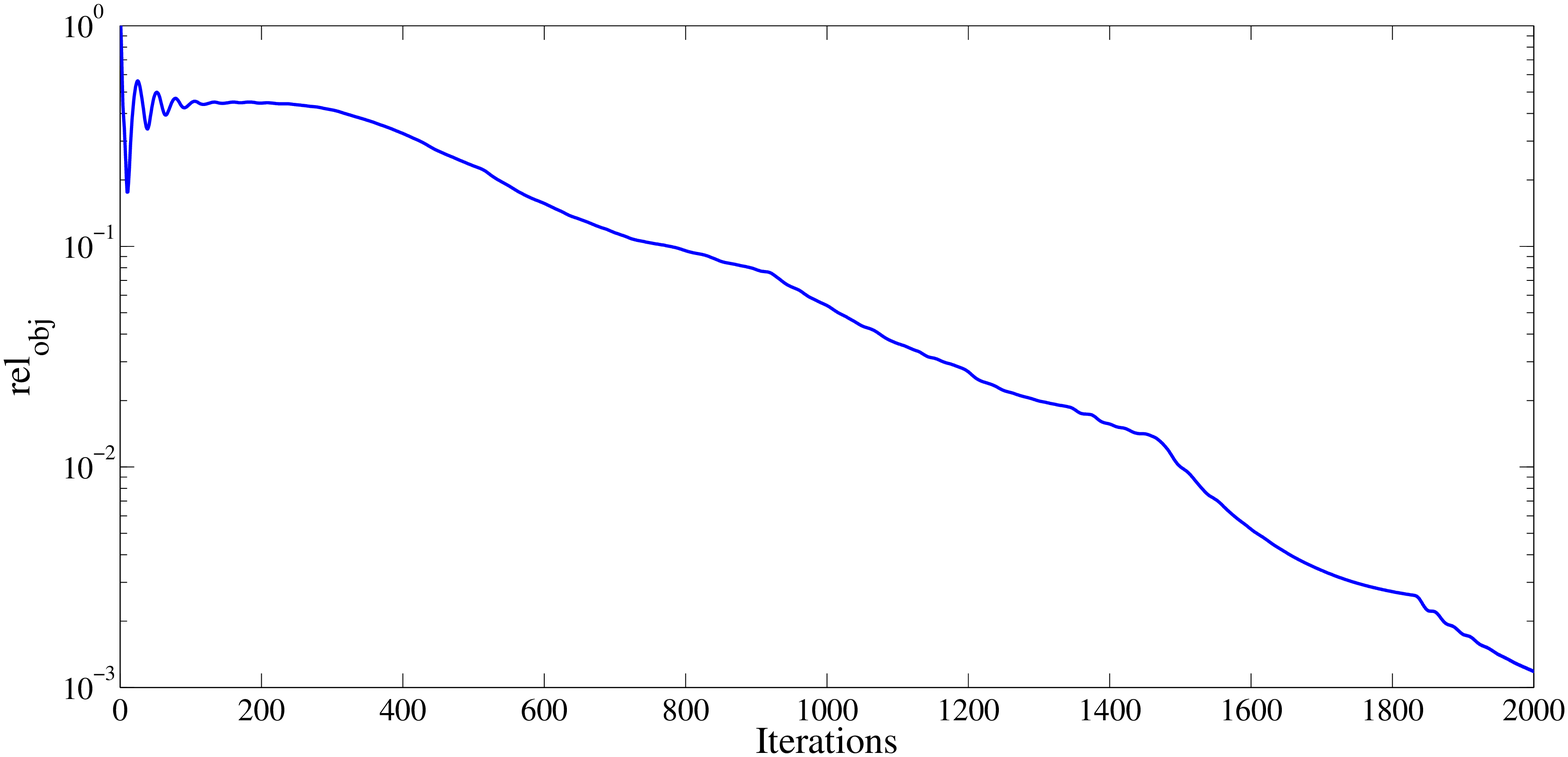}
 \caption{Relative distance from objective function.}
 \label{rel-obj}
 \vspace{.3cm}
\end{figure}

\begin{figure}[t]
 \centering
\includegraphics[width=0.5\textwidth]{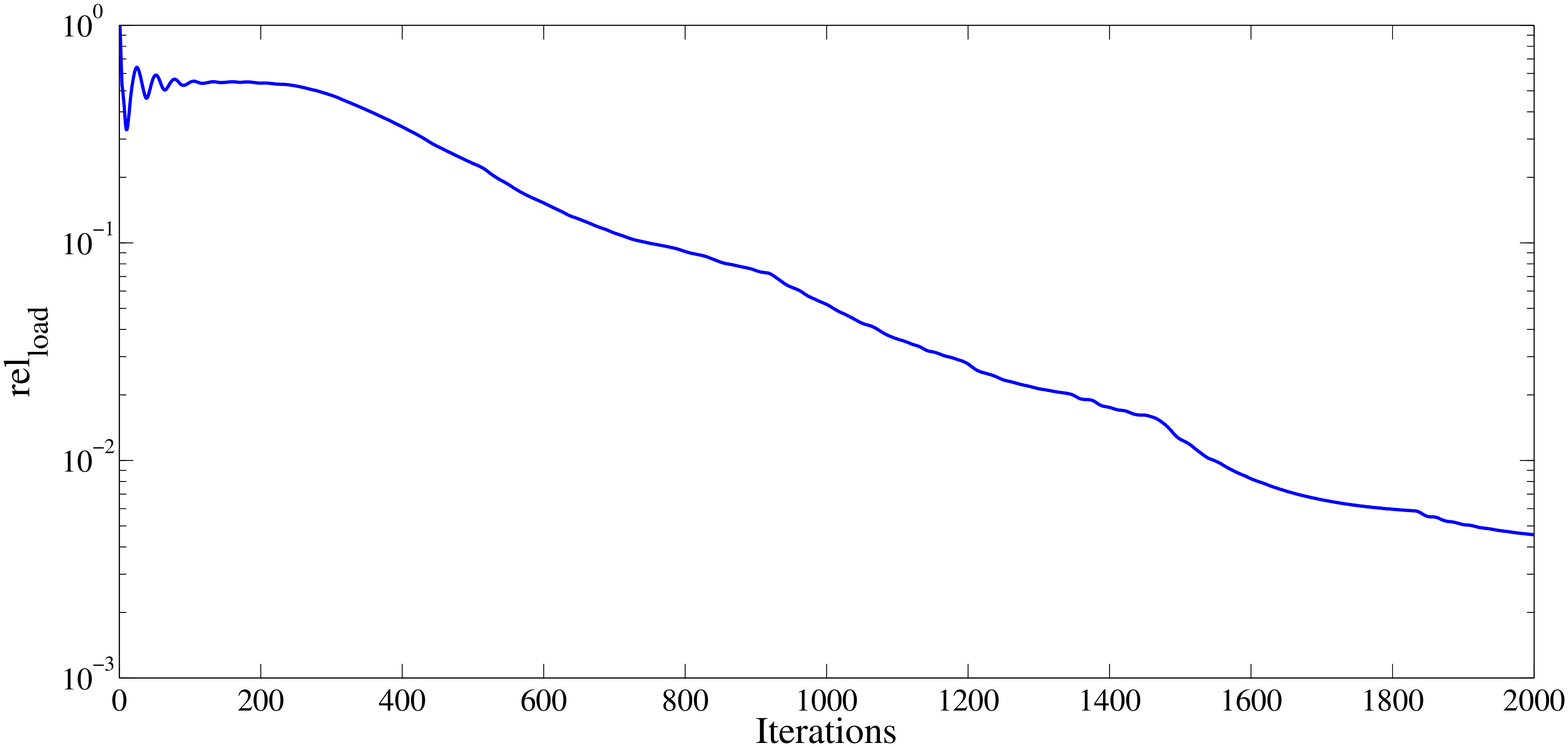}
 \caption{Relative distance from optimal load.}
 \label{rel-load}
\end{figure}

Figure~\ref{EVcharging} shows the evolution of the total daily charging load for 10 selected PEVs over 2000 iterations. Note, the oscillations could be prevented by reducing some of the tuning parameters, although this might lead to a larger number of iterations until convergence.  Moreover, Figs.~\ref{rel-obj} and~\ref{rel-load} illustrate the two introduced convergence measures, i.e., $\mathrm{rel}_\mathrm{obj}$ and $\mathrm{rel}_\mathrm{load}$, over the course of iterations with the values being 0.0028 and 0.0056 after 2000 iterations.

Note, each iteration of $\mathcal{CI-PEVCC}$ is computationally inexpensive since it only requires evaluation of algebraic functions which could be done in parallel. In the above figures, each iteration only corresponds to variables updates according to updates \eqref{LambdaUpdate}-\eqref{MUupdate}. The depicted intermediate values do not necessarily constitute a feasible solution for the PEV-CC problem.

By adding only one more communication link, and turn the communication topology into a ring, the $res_{obj}$ and $res_{obj}$ after 2000 iterations decrease to 0.0012 and 0.0046, respectively. Note, adding this extra communication link decreases the diameter of communication graph approximately by half, hence, increases the speed of information spread across the agents. Therefore, the convergence of algorithm improves. Adding more communication links could potentially further improve the convergence speed (see \cite{mohammadi2014role}).

The computer simulations are executed using MATLAB on a PC with a Core i-7 processor (2.7 GHz) and 8 GB RAM, and the average combined CPU time for all agents per
iteration is 0.004 s. It is worth mentioning that by assigning the computational load to distributed agents in real-world applications, the computations could be executed in parallel at each iteration.

\section{Conclusion}
In this paper, we have proposed a fully distributed \textit{consensus+innovation}-based approach to solve the PEV's coordinated charging problem, i.e.,  charging schedules are determined such that the cost to supply demand is minimized while each PEV's constraints are fulfilled. The main features of the algorithm are that
it enables a fully distributed implementation down to the PEV level without the need for a coordinator. Each PEV has to update/evaluate simple functions over the course of iterations while information exchange is restricted to communicating a Lagrange multiplier, which defines the value of consumption from each PEV point of view, with neighboring agents. In particular, there is no need to share information about driving patterns or charging schedules. Also, the communication graph could be defined arbitrarily as long as it is connected. Moreover, our distributed algorithm could easily capture individual cost functions for the PEVs, e.g., battery degradation costs and drivers utility as a function of battery's SOC. The algorithm has been tested on a fleet of 100 PEVs showing that it converges to the overall optimal solution.

\section*{Acknowledgment}
This work was supported in part by the National Science Foundation under grant ECCS-1408222. Also, we would like to acknowledge the  inputs of  Prof.  G\"oran Andersson to this paper.

\bibliographystyle{IEEEtran}
\bibliography{References1,ReferencesAreaOPF,ReferencesDSEV,literatur}

\vfill

\end{document}